\documentclass[twocolumn,superscriptaddress,amsmath,aps,nofootinbib,nolongbibliography]{revtex4-2}
\usepackage[version=3]{mhchem}
\usepackage{hyperref}
\usepackage{graphicx}
\usepackage{amsmath,amsfonts}
\usepackage{dcolumn}
\usepackage{bm}
\usepackage[dvipsnames]{xcolor}
\usepackage{multirow}
\usepackage{physics}
\usepackage{subfigure}
\usepackage[utf8]{inputenc}
\usepackage{titlesec}
\usepackage{float}
\usepackage[margin=0.6in]{geometry}

\hypersetup{
  colorlinks=true,
  linkcolor=blue,
  citecolor=blue,
  urlcolor=blue
}

\begin{document}

\title{Anisotropic magnetism of polymorphic \ce{ErAl3}}

\author{Karolina Gornicka}
\affiliation{Materials Science and Technology Division, Oak Ridge National Laboratory, Oak Ridge, Tennessee 37831, United States}
\affiliation{Faculty of Applied Physics and Mathematics and Advanced Materials Centre, Gdansk University of Technology, ul. Narutowicza 11/12, 80-233 Gdańsk, Poland}

\author{Brenden R. Ortiz}
\affiliation{Materials Science and Technology Division, Oak Ridge National Laboratory, Oak Ridge, Tennessee 37831, United States}

\author{Andrew D. Christianson}
\affiliation{Materials Science and Technology Division, Oak Ridge National Laboratory, Oak Ridge, Tennessee 37831, United States}

\author{Andrew F. May}
\email{mayaf@ornl.gov}
\affiliation{Materials Science and Technology Division, Oak Ridge National Laboratory, Oak Ridge, Tennessee 37831, United States}

\date{\today}

\begin{abstract}
ErAl$_3$ can form in either a trigonal ($\alpha$) or cubic ($\beta$) polymorph and this paper investigates the physical properties of these polymorphs through characterizations of single crystals grown in an aluminum flux. We demonstrate that polymorph selection can be achieved based on the nominal composition of the crystal growth.  Magnetic measurements confirm that both $\beta$-ErAl$_3$ and $\alpha$-ErAl$_3$ order antiferromagnetically at low temperatures. $\beta$-ErAl$_3$ undergoes AFM ordering at a Néel temperature \textit{T$_N$} = 5.1 K, and the transition is suppressed continually with applied field. $\alpha$-ErAl$_3$ displays more complex behavior, with successive magnetic transitions at \textit{T$_N$} = 5.7 K and \textit{T$_2$} = 4.6 K for zero field, where heat capacity and dilatometry measurements evidence that these transitions are second- and first-order, respectively. Under magnetic field, strong anisotropy is revealed in $\alpha$-ErAl$_3$, with several step-like metamagnetic transitions observed below \textit{T$_2$} for \textit{H}$\parallel$c. These transitions produce sequential magnetization plateaus near one-half of the apparent saturation magnetization. The electrical resistivity of $\alpha$-ErAl$_3$ is strongly coupled to its magnetism. At \textit{T} = 2 K, we observe a positive magnetoresistance reaching 60\%, with distinct anomalies at the metamagnetic transitions. The results are summarized in \textit{H} - \textit{T} phase diagrams that demonstrate complex magnetic behavior for $\alpha$-ErAl$_3$, suggesting an important role of competing interactions in this metallic system that possesses characteristics of Ising physics.
\end{abstract}

\maketitle

\section{Introduction}
The \textit{R}Al$_3$ family, where \textit{R} = rare earth metal, has raised a lot of interest in recent years due to  complex magnetism, heavy-fermion behavior, valence fluctuations, and the crucial role of the crystal electric field (CEF) in determining their electronic and magnetic properties \cite{cornelius2002two,blyth1993temperature,yamamoto2022magnetic,coles1987frustration,buschow1965system,mader1968magnetic,weber1977magnetic,strassle2003crystal,will1974magnetic,bargouth1971magnetic,buschow1968crystalline,amin2017structural,bud2007thermal,de1970nuclear,forker2007electric,sugiyama2001high,ahmed2024large,buschow1966magnetic,ebihara2000heavy, hiess2000magnetism, deutz1989magnetic}. The rare earth elements play a critical role in shaping the behavior of these compounds, and most of them exhibit long-range magnetic order at low temperatures \cite{buschow1965system,coles1987frustration,yamamoto2022magnetic,ahmed2024large}. GdAl$_3$, for instance, is a frustrated magnetic material that orders antiferromagnetically at 17 K, as confirmed by electron spin resonance spectroscopy and magnetization measurements \cite{coles1987frustration,buschow1966magnetic}.

The size of the rare earth atom has a strong influence on the crystal structure of the \textit{R}Al$_3$ materials \cite{cannon1975effect,van1966structures, gao2007lattice}. As the rare-earth atomic radius decreases, the \textit{R}Al$_3$ equilibrium structure changes from a hexagonal lattice (SnNi$_3$ - type, also known as Mg$_3$Cd - type) for \textit{R} = La - Gd, to the cubic AuCu$_3$ structure type for \textit{R} = Tm, Yb, and Lu. For intermediate atomic sizes, such as Er, Dy or Ho, two phases of \textit{R}Al$_3$ may form, i.e. the cubic $\beta$ phase (AuCu$_3$-type structure) and the trigonal $\alpha$ phase (HoAl$_3$-type structure) \cite{meyer1970kubische, xu2022experimental}. For \textit{R} = Er, Meyer \cite{meyer1970kubische} reported that the cubic $\beta$-ErAl$_3$ is the only stable phase in uncontaminated ErAl$_3$. However, Meyer also noted that the trigonal $\alpha$-ErAl$_3$ phase can be stabilized by the presence of a small amount of Si impurity. On the other hand, Gao et al. \cite{gao2021phase} investigated the thermodynamics of ErAl$_3$ using  first-principles calculations and experimental analysis. They reported that the HoAl$_3$-type structure is the thermodynamically stable phase at ambient condition, while the AuCu$_3$-type structure of ErAl$_3$ is a metastable phase.  In particular, they observed that the cubic $\beta$-ErAl$_3$ phase formed at a cooling rate of 60 $^\circ$C/h, whereas the trigonal $\alpha$-ErAl$_3$ emerged at a much lower cooling rate of 5 $^\circ$C/h during the solidification of an Er$_{1.1}$Al$_3$ melt. This may suggest that the cubic phase forms at high temperature and can become kinetically trapped (metastable). Such behavior appears consistent with that of HoAl$_3$, where rapid quenching locks in the metastable cubic phase while slow cooling produces the trigonal structure \cite{yamamoto2022magnetic}.

\begin{figure}[b]
\centering
\includegraphics[width=0.49\textwidth]{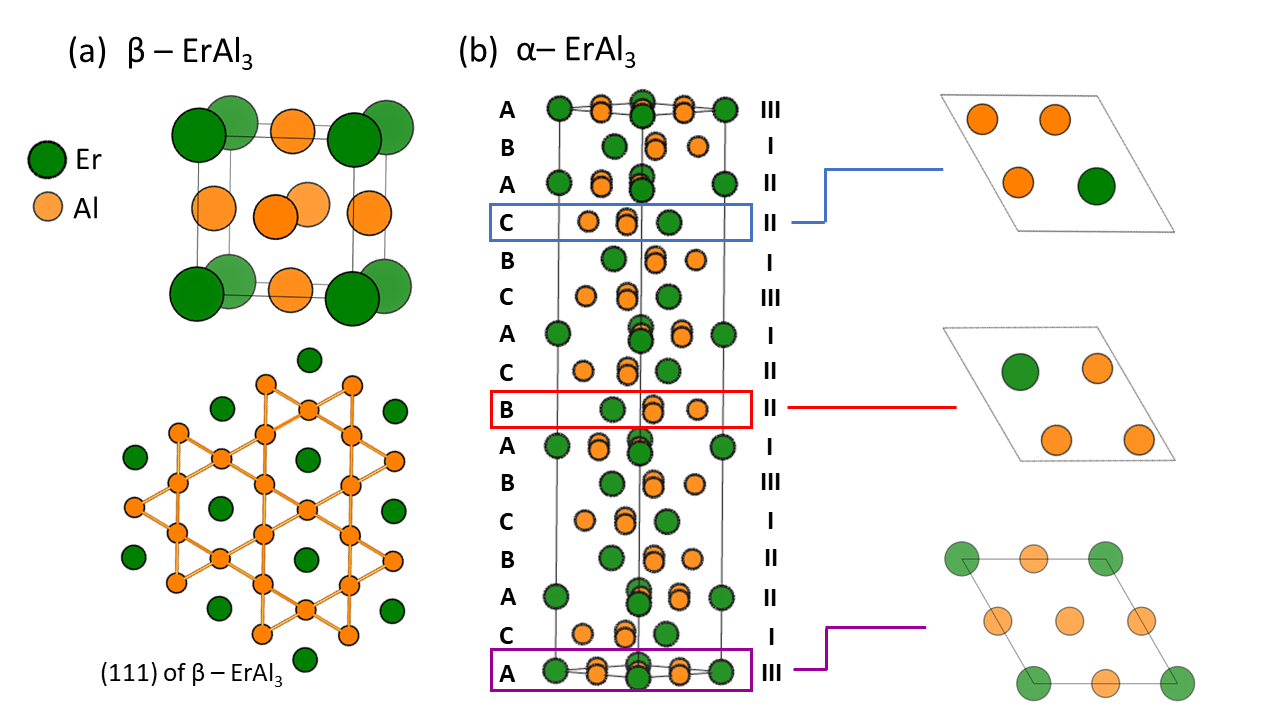}
\caption{Crystal structures of the  (a) cubic $\beta$-ErAl$_3$ and (b) trigonal $\alpha$-ErAl$_3$.  The structure of $\alpha$-ErAl$_3$ is built from the (111) planes of $\beta$-ErAl$_3$ in the stacking sequence indicated by letters A,B,C. The layer types I and II have a slight breathing-mode distortion of the Al-derived kagome net, while layer type III retains an the ideal kagome motif as in the (111) plane of $\beta$-ErAl$_3$.}
\label{fig:crystal}
\end{figure}

As mentioned above, ErAl$_3$ can crystallize in two distinct structures: cubic (\textit{$Pm\bar{3}m$}) and trigonal (\textit{$R\bar{3}m$}). In the cubic structure (see Fig.\ref{fig:crystal}a), each atom is surrounded by twelve neighbors, highlighting a close-packed arrangement. On the other hand, the crystal structure of trigonal $\alpha$-ErAl$_3$ consists of close-packed layers stacked along the \textit{c}-axis as presented in Fig.\ref{fig:crystal}b.  These layers are derived from the (111) plane of the cubic $\beta$-ErAl$_3$, and their stacking sequence is marked by letters A, B, or C in Fig.\ref{fig:crystal}b.  The stacking sequence indicates a shifting of layers within the $ab$-plane while translating along [001], and the full unit cell of $\alpha$-ErAl$_3$ has a rather complicated stacking sequence. Separately, there are three crystallographically distinct types of layers, denoted as I, II, and III. Layer III has the undistorted kagome network of Al atoms found in the (111) plane of the cubic structure, while layers I and II have a breathing-mode distortion that leads to larger and smaller Al-Al triangles within the layer. In each layer, the Er atoms form a triangular net with the edge equal to the cell length \textit{a} of 0.60 nm, and this geometry may promote magnetic frustration. However, the interlayer Er-Er distances of 0.42 nm are significantly shorter than the intralayer ones, so physical properties are expected to be three-dimensional. Importantly, rare earth intermetallics can host sizable further neighbor interactions mediated by the conduction electrons (Ruderman-Kittel-Kasuya-Yosida (RKKY) mechanism \cite{ruderman1954indirect,kasuya1956theory,van1962note}), which can produce complex quantum ground states\cite{manuel1999magnetic,iqbal2016spin,mishmash2013theory,chang2024exploring}. Such interactions can combine with anisotropic terms to stabilize complex spin structures in applied fields, and the results can vary significantly between Ising systems and those with negligigle orbital anisotropy.
 
Interestingly, there are significant inconsistencies in the existing literature on the physical properties of $\alpha$-ErAl$_3$ and $\beta$-ErAl$_3$ compounds \cite{ahmed2024large,bargouth1971magnetic,buschow1968crystalline,buschow1966magnetic,will1974magnetic}. The magnetic structure of $\alpha$-ErAl$_3$ was first determined through magnetization and neutron powder diffraction experiments using polycrystalline samples \cite{buschow1968crystalline,bargouth1971magnetic}. It was found that $\alpha$-ErAl$_3$ orders antiferromagnetically somewhat below 5 K with the moments directed along the trigonal \textit{c}-axis\cite{buschow1968crystalline}. On the other hand, Ahmed et al. \cite{ahmed2024large} recently reported three transition temperatures (4.67, 5.63, and 12.87 K) for a polycrystalline sample, claiming that a paramagnetic-ferromagnetic transition exists at 12.87 K. In the case of $\beta$-ErAl$_3$, previous reports of magnetic ordering are also highly inconsistent: one study observes ferromagnetic behavior at 21 K \cite{buschow1966magnetic}, while another study identifies antiferromagnetic ordering below 5 K \cite{will1974magnetic,buschow1968crystalline}. This divergence in findings motivated us to undertake investigations using single crystals to re-evaluate and determine the intrinsic magnetic properties of $\beta$-ErAl$_3$ and $\alpha$-ErAl$_3$. In addition, the previous data were reported on polycrystalline samples and thus the magnetic anisotropy was not investigated. 

In this work, we grew single crystals of cubic $\beta$-ErAl$_3$ and trigonal $\alpha$-ErAl$_3$, and here report their magnetic, thermal, and electronic transport properties.  Polymorph selection is achieved based on the nominal composition of the melt, suggesting temperature-dependent phase stability for $\alpha$-ErAl$_3$ and $\beta$-ErAl$_3$.  Our studies of single crystals show that $\beta$-ErAl$_3$ orders antiferromagnetically below 5.1 K. The transition is suppressed continuously with increasing field, consistent with what is observed for other AFM systems where an applied field leads to the gradual polarization of spins.  $\alpha$-ErAl$_3$, on the other hand, displays complex magnetic behavior at low temperatures. Antiferromagnetic ordering appears below \textit{T$_N$} = 5.7 K with an additional transition at \textit{T$_2$} = 4.6 K, which are second- and first-order transitions, respectively. Applying a magnetic field parallel to the \textit{c}-axis induces metamagnetic transitions with magnetization plateaus near one-half of the saturation moment for $T$ $<$ $T_2$, and corresponding features are observed in the magnetotransport. The rich \textit{H}-\textit{T} phase diagram suggests that $\alpha$-ErAl$_3$ is a promising system for studying Ising or strongly-anisotropic behaviors in rare earth antiferromagnets with three-dimensional character.

\section{Experimental details}

\subsection{Single crystal growth}

 Single crystals of $\alpha$-ErAl$_3$ and $\beta$-ErAl$_3$ were grown using an Al self-flux method. High purity  Er (RE products, 4N) and Al (Thermo Scientific, 4N) were weighed to a molar ratio of 5:95 and 10:90 for $\alpha$-ErAl$_3$ and $\beta$-ErAl$_3$, respectively, placed into 5 mL Canfield crucibles fitted with a catch crucible and a porous frit \cite{canfield2016use}, and sealed in fused silica ampoules under a partial pressure of Ar. Samples were subsequently heated in a resistance furnace up to 1100$^\circ$C and held at this temperature for 12 hours. The furnace was then cooled down to 700$^\circ$C at a rate of 3$^\circ$C per hour, and the crystals were isolated by decanting the excess flux using a centrifuge. The cooling rate of 3$^\circ$C/hr, utilized for both the cubic $\beta$-ErAl$_3$ and trigonal $\alpha$-ErAl$_3$ phases, was sufficient to obtain large, stable single crystals.  The obtained crystals were shiny and found to be stable in air over a period of months.

\subsection{Characterization}

Room temperature powder X-ray diffraction (pXRD) measurements were carried out on ground crystals using a PANalytical X’pert Pro diffractometer equipped with monochromated Cu K$\alpha_1$ radiation. Crystal structure analysis was performed by LeBail refinements using the TOPAS software. Single crystal X-ray diffraction (SCXRD) data were collected on a Bruker D8 Advance Quest diffractometer with a graphite monochromator using Mo K$\alpha$ ($\lambda$=0.71073 \AA). Single crystals were mounted on a Kapton loop and glued with glycerol. The structural solutions were obtained and refined by the full-matrix least-squares method with the Bruker APEX4 software package. The crystal structure was visualized with VESTA software. Elemental analysis was carried out on as-grown crystals using a Hitachi-TM3000 microscope equipped with a Bruker Quantax 70 energy dispersive spectroscopy (EDS) system.

Magnetization measurements were carried out using a 7 T Quantum Design (QD) Magnetic Property Measurement System (MPMS3) SQUID magnetometer in the temperature range 2 - 300 K. Magnetization data were collected while cooling in an applied field. For the trigonal $\alpha$-ErAl$_3$ the anisotropic magnetization was measured having the external field parallel to the \textit{c}-axis (\textit{H}$\parallel$c) or applied within the \textit{ab}-plane (\textit{H}$\parallel$ab). For the cubic $\beta$-ErAl$_3$, magnetic measurements were performed on ground crystals and with the magnetic field parallel to the [100]. Several magnetization vs field (\textit{M}-\textit{H}) isotherms in fields up to 70 kOe were measured at different temperatures. Additional \textit{M}-\textit{H} measurements on $\alpha$-ErAl$_3$ at $T$ = 0.4 K utilized the Quantum Design iHe-3 $^3$He insert for the MPMS3. A field-dependent magnetization measurement for $\alpha$-ErAl$_3$ (\textit{H}$\parallel$c, \textit{T} = 2 K) up to 135 kOe was performed using the QD PPMS 14T with the ACMS-II option.

Electrical resistivity measurements were conducted in a Quantum Design 9 T PPMS in the temperature range 2 - 300 K using a standard four-probe technique. Platinum electrical leads were attached to the sample using silver paint (DuPont 4929N) and the sample was attached to a sapphire plate with GE varnish. Heat capacity measurements were performed using the heat-capacity option on the same PPMS. Heat capacity was measured using both a standard 2\% temperature rise and the large heat pulse method with a 30\% rise, with an appropriate addenda. For the trigonal $\alpha$-ErAl$_3$ dilatometry experiments were performed using Quantum Design’s fused silica capacitive cell. The dilatometer was mounted in a Quantum Design Dynacool and was operated over a temperature range 380 - 2 K. The thermal expansion was measured along the \textit{c}-axis.

\begin{figure}[t]
\centering
\includegraphics[width=0.95\columnwidth]{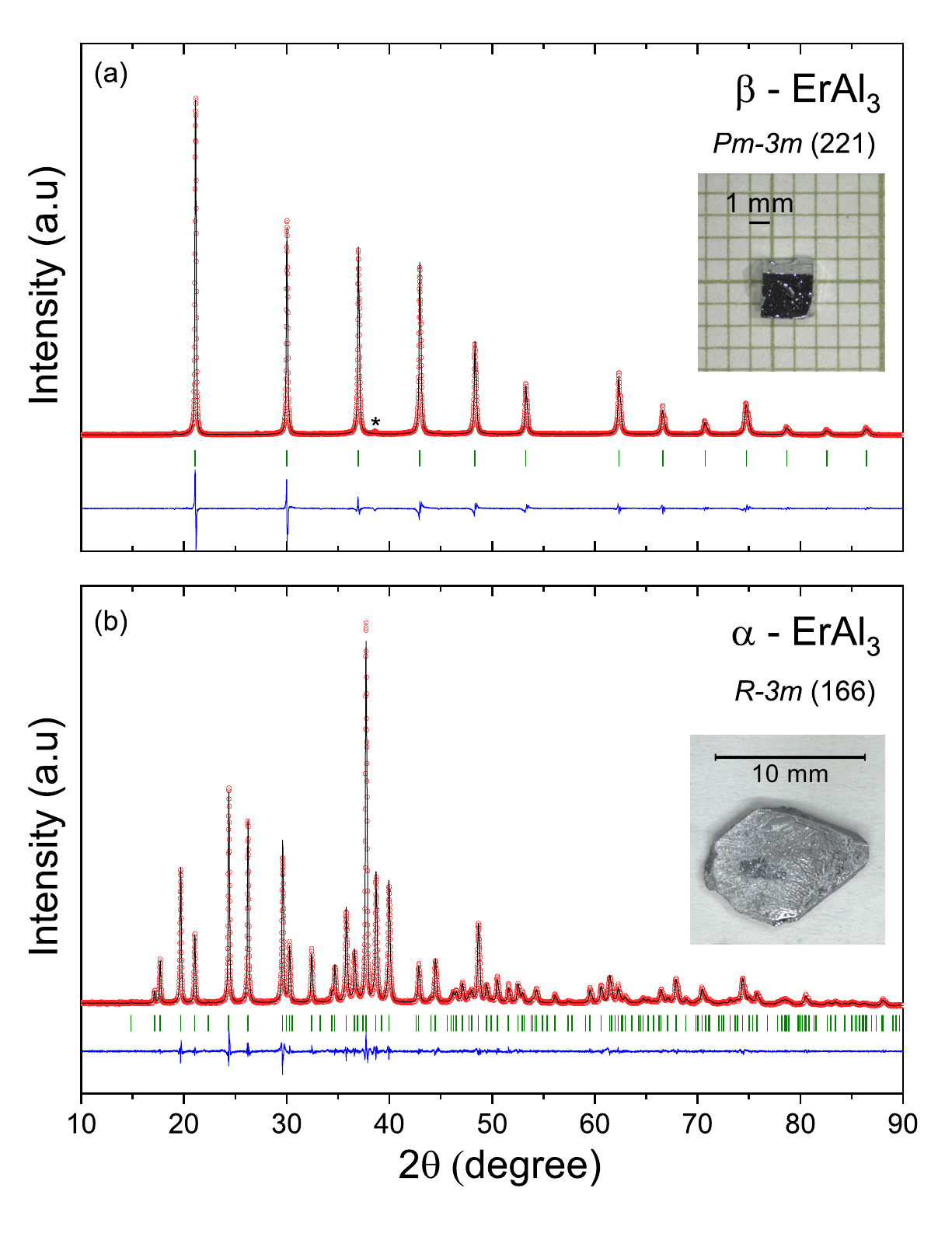}
\caption{Powder XRD patterns of the crushed (a) $\beta$-ErAl$_3$  and (b) $\alpha$-ErAl$_3$ single crystals with the LeBail refinement results. The weak peak marked with the star (*) results from the aluminum flux. The red dots are the experimentally observed data, and the black line is the calculated XRD pattern. The green vertical bars represent the allowed Bragg positions. The blue solid line represents the difference between the experimental and calculated diffraction patterns. Insets show photos of single crystals.}
\label{fig:pxrd}
\end{figure}

\section{Results and discussion}

\subsection{Crystal structure}

\begin{figure*}[ht!]
\centering
\includegraphics[width=0.9\textwidth]{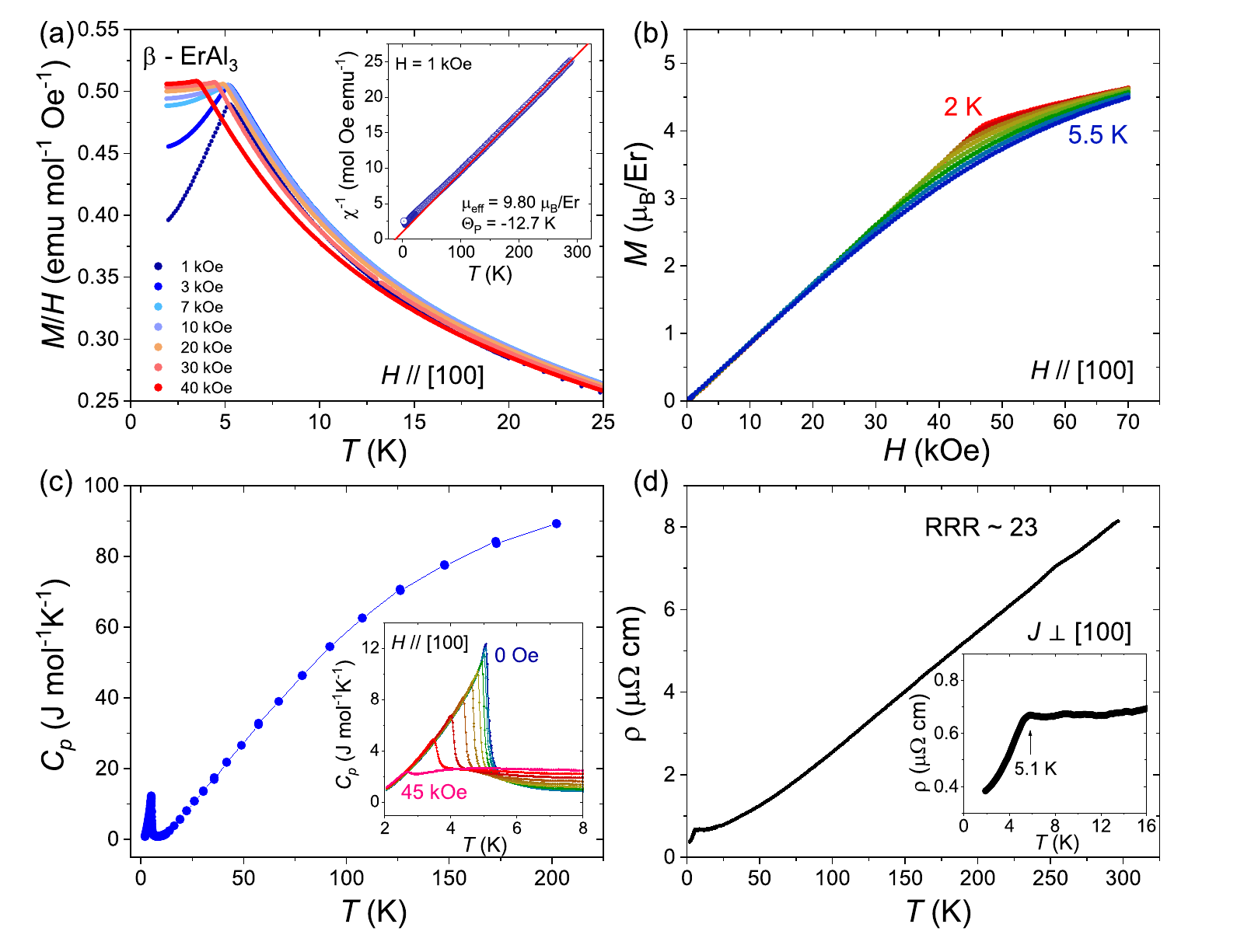}
\caption{Physical property data for single crystalline cubic $\beta$-ErAl$_3$. (a) Temperature dependence of \textit{M/\textit{H}} and inverse magnetic susceptibility (inset) of a $\beta$-ErAl$_3$ single crystal measured with magnetic fields parallel to [100]. (b) Magnetization as a function of magnetic field measured from 2 K to 5.5 K with $\Delta$\textit{T} = 0.5 K. (c) The temperature dependence of heat capacity in the absence of external magnetic field. The inset shows the low-temperature anomaly evolve with magnetic field ($\Delta$\textit{H} = 5 kOe). (d) Temperature dependence of resistivity with inset depicting an enlarged view near the phase transition, where the vertical arrow indicates $T_N$ determined from magnetization data.}
\label{fig:beta}
\end{figure*}

X-ray diffraction was used to confirm phase purity and crystal structure of the single crystals. As can be seen in the insets in Figs.\ref{fig:pxrd}a and \ref{fig:pxrd}b, the grown single crystals were up to a few mm in size and have well developed shapes corresponding to the lattice symmetry. Our results indicate that the cubic AuCu$_3$-type structure and the trigonal HoAl$_3$-type phase form depending on the initial Er:Al molar ratio, without the need for contamination. Generically, the starting molar ratio impacts the temperature and composition at which crystallization can occur while cooling. Thus, the selection of different polymorphs is likely driven by the temperature at which the crystal grows, with different polymorphs thermodynamically stable at different temperatures. The potential for a chemical (compositional) component to this stability cannot be ruled out at this time, but our EDS was not able to detect a difference in composition between these different ErAl$_3$ phases.

Single crystal x-ray diffraction experiments confirmed the crystal structure and good crystalline quality for both phases of ErAl$_3$. The resulting CIF files are included in the Supplementary Information\cite{Supplemental}. The space group and lattice parameters obtained were taken as the starting point for the LeBail fit of the powder X-ray diffraction patterns collected on crushed $\beta$-ErAl$_3$ and $\alpha$-ErAl$_3$ crystals (see Fig. \ref{fig:pxrd}). The LeBail method was preferred over the Rietveld method due to the texturing in the powder samples. The presented patterns demonstrate the two phases of ErAl$_3$, and sharp peaks indicate high crystalline quality of the grown single crystals. The LeBail refinement confirms that $\beta$-ErAl$_3$ crystallizes in the AuCu$_3$-type structure with the cubic space group \textit{$Pm\bar{3}m$} (221), and the pXRD pattern of $\alpha$-ErAl$_3$ can be entirely indexed by the expected trigonal \textit{$R\bar{3}m$} (166) space group. The corresponding lattice parameters \textit{a} = 4.2082(2) Å for $\beta$-ErAl$_3$, and \textit{a} = 6.03219(4) Å and \textit{c} = 35.7517(3) Å for $\alpha$-ErAl$_3$ are in good agreement with previously reported values for polycrystalline samples \cite{buschow1965system,buschow1967systematic,gschneidner1988er,van1966structures,moriarty1965some, ahmed2024large}. We also observed very weak Bragg peaks from the aluminum flux utilized (marked with a star in Fig.\ref{fig:pxrd}a); since this is a non-magnetic impurity, it does not impact the results in this study.

\subsection{Physical properties of $\beta$-ErAl$_3$}
In this section, we discuss the physical properties of cubic $\beta$-ErAl$_3$. Magnetic, thermodynamic and electronic properties of $\beta$-ErAl$_3$ single crystals are summarized in Fig. \ref{fig:beta}. We also performed magnetization measurements on powdered crystals (not shown here) and observed that the results are identical to those obtained from single crystal measurements, implying negligible magnetic anisotropy in $\beta$-ErAl$_3$. 

The temperature dependence of the field-cooled magnetization measured with the magnetic field applied parallel to the [100] is shown in Fig.\ref{fig:beta}a.  At low temperatures, the magnetization exhibits a cusp-like maximum for \textit{H} = 1 kOe, indicating the formation of antiferromagnetic order upon cooling. The magnetic ordering temperature determined from \textit{d}(\textit{MT})/\textit{dT} is \textit{T$_N$} = 5.1 K. The cusp near \textit{T$_N$} shifts to lower temperatures with increasing external field, which is consistent with AFM order. Our results are consistent with magnetization and neutron diffraction measurements performed on a polycrystalline sample \cite{will1974magnetic,buschow1968crystalline}, thereby ruling out the possibility of a ferromagnetic transition at 21 K as suggested in Ref. \cite{buschow1966magnetic}.

The inset of Fig.\ref{fig:beta}a presents the temperature dependence of the inverse magnetic susceptibility ($\chi^{-1}$), where $\chi$ is defined as \textit{M}/\textit{H}. The data in the paramagnetic (PM) state are well described by the Curie-Weiss law, $\chi$(\textit{T}) = \textit{C}/(\textit{T} - $\theta_p$), where \textit{C} is the Curie constant and $\theta_p$ is the Weiss temperature. A linear fit of $\chi^{-1}$(\textit{T}) in the temperature range 150 - 300 K is shown by the straight red line. The least-square fitting yields the value of $\theta_p$ = -12.7(3) K, suggesting a predominance of antiferromagnetic exchange interactions in the paramagnetic state of $\beta$-ErAl$_3$, though this determination may be impacted by the crystal field splitting of the Er$^{3+}$ manifold. The effective moment obtained from the fitted value of the Curie constant is $\mu_{eff}$ = 9.80(2) $\mu_B$/Er, which is close to the theoretical value for the Er$^{3+}$ free ion moment (\textit{g}\(\sqrt{\textit{S}(\textit{S}+1)}\) = 9.58 $\mu_B$, \textit{S} = \(\frac{3}{2}\) and \textit{g} = \(\frac{6}{5}\)). Note that an estimation of $\mu_{eff}$ is not available from earlier work on a polycrystalline sample \cite{buschow1968crystalline}.

We also measured isothermal magnetization for \textit{H}$\parallel$[100] between 2 and 5.5 K ($\Delta$\textit{T} = 0.5 K) with magnetic fields up to 70 kOe, as presented in Fig.\ref{fig:beta}b. The \textit{M} - \textit{H} curves possess simple behavior and increase monotonically with the magnetic field reaching a value of 4.65 $\mu_B$/Er ($T$ = 2 K, $H$ = 70 kOe). A small kink can be observed at the lowest temperatures, corresponding to suppression of antiferromagnetic order and apparent saturation. The observed value of 4.65 $\mu_B$ is in good agreement with the magnetic moment determined from M{\"o}ssbauer spectroscopy \cite{zinn1968mossbauer} as well as the neutron diffraction data obtained for the polycrystalline sample \cite{will1974magnetic}. If this value is compared with 9 $\mu_B$, the saturation moment of the free Er$^{3+}$ ion, it becomes evident that the value is reduced, likely due to strong crystal field effects (CEF).

The temperature dependence of the zero-field heat capacity ($\textit{C}_P$) for $\beta$-ErAl$_3$ is shown in Fig.\ref{fig:beta}c. The bulk nature of the magnetic transition is confirmed by the $\lambda$-shape peak with appreciable magnitude near 5 K. The evolution of the peak under selected applied magnetic fields is presented in the inset of Fig.\ref{fig:beta}c. Without the magnetic field, a sharp peak is observed at \textit{T$_N$} = 5.05 K, in very good agreement with the magnetization data. When magnetic fields are applied, this peak is suppressed in magnitude, becomes broader, and shifts towards lower temperatures, consistent with long-range AFM ordering. 

Figure \ref{fig:beta}d contains the temperature-dependent electrical resistivity ($\rho$) of a $\beta$-ErAl$_3$ single crystal. The $\rho$(\textit{T}) of $\beta$-ErAl$_3$ has metallic behavior down to 2 K with a sharp anomaly at 5.1 K. The drop in resistivity (see inset in Fig.\ref{fig:beta}d) is associated with the magnetic phase transition where $\rho$(\textit{T}) decreases rapidly, presumably due to a reduction of spin scattering as the system orders magnetically. The residual resistivity ratio [RRR = $\rho$(300 K)/$\rho$(2 K)] $\approx$ 23 suggests the single crystals are of good quality. The minor feature observed near 250 K is likely due to a measurement artifact.

As it has been shown, our measurements performed on single crystals confirm that the cubic $\beta$-ErAl$_3$ exhibits a distinct antiferromagnetic transition at 5.1 K. Even though our results differ from an earlier study, which reported a ferromagnetic transition at 21 K \cite{buschow1966magnetic}, they are consistent with more recent investigations that observe AFM ordering below 5 K \cite{buschow1968crystalline,will1974magnetic}.

\subsection{Physical properties of $\alpha$-ErAl$_3$}
 
\begin{figure}[t]
\centering
\includegraphics[width=0.49\textwidth]{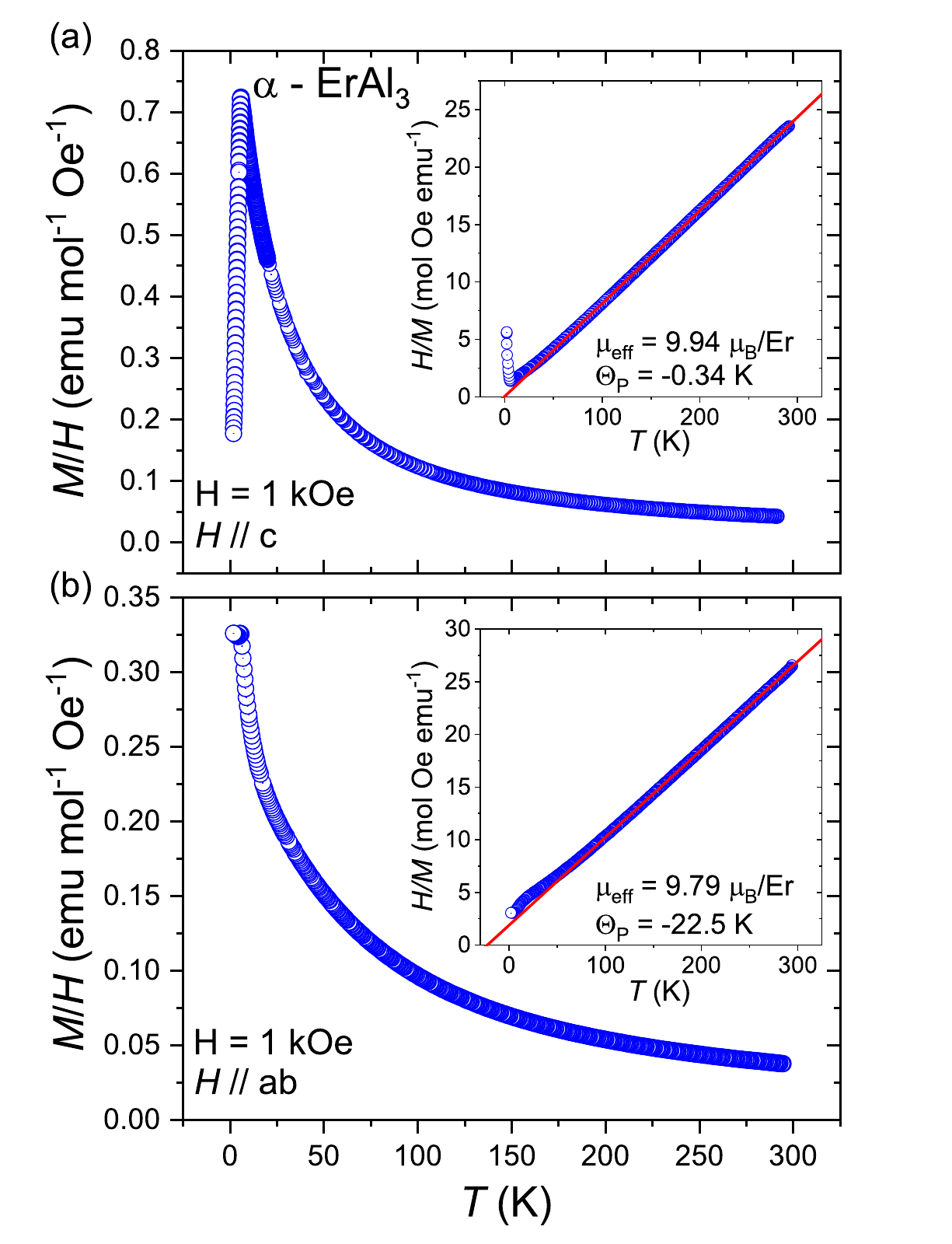}
\caption{Temperature dependence of magnetic susceptibility of $\alpha$-ErAl$_3$ crystals with the magnetic field (1 kOe) applied (a) parallel and (b) perpendicular to the \textit{c}-axis. Insets show inverse magnetic susceptibility together with the Curie-Weiss fit performed above 100 K.}
\label{fig:alpha_MvsT_1000Oe}
\end{figure}

We now consider the anisotropic magnetization of the trigonal phase of ErAl$_3$. Fig.\ref{fig:alpha_MvsT_1000Oe}a and \ref{fig:alpha_MvsT_1000Oe}b contain the FC magnetization curves of a $\alpha$-ErAl$_3$ single crystal measured in \textit{H} = 1 kOe aligned along the \textit{c}-axis (\textit{H}$\parallel$c) and in the \textit{ab}-plane (\textit{H}$\parallel$ab), respectively. When \textit{H}$\parallel$c, a sharp cusp is observed at 5.7 K in the \textit{M}(\textit{T}) data. This indicates an AFM ordering and marks the boundary between the AFM and PM phases. As shown in Fig.\ref{fig:alpha_MvsT_1000Oe}, magnetization measurements reveal significant magnetic anisotropy in $\alpha$-ErAl$_3$. When the magnetic field is applied parallel to the \textit{c}-axis, the magnetization decreases sharply with decreasing temperature below \textit{T$_N$}. In contrast, when the field is aligned along the \textit{ab}-plane, the magnetization is nearly temperature-independent below \textit{T$_N$}. These findings indicate that the ordered moments are oriented along the \textit{c}-axis in the ground state, in agreement with previous neutron diffraction data\cite{bargouth1971magnetic}. It should be noted that we do not observe a paramagnetic-ferromagnetic (FM) transition around  13 K, as was reported recently for a polycrystalline sample \cite{ahmed2024large}. The purported PM-FM transition temperature of 12.87 K \cite{ahmed2024large} is very close to the Curie temperature reported for ErAl$_2$ (12.6 K \cite{del1987anisotropy,jaakkola1980dependence}). It is quite possible that ErAl$_2$ might form as a secondary phase in a polycrystalline sample and affect the observed magnetic properties. 

The inverse magnetic susceptibility as a function of temperature for \textit{H}$\parallel$c and \textit{H}$\parallel$ab are shown in insets of Fig.\ref{fig:alpha_MvsT_1000Oe}a and \ref{fig:alpha_MvsT_1000Oe}b, respectively. For both field orientations, the data above 100 K can be well fitted by the Curie-Weiss formula. The effective moments in the PM state were found to be $\mu_{eff}$ = 9.94(1) $\mu_B$/Er for \textit{H}$\parallel$c and $\mu_{eff}$ = 9.79(3) $\mu_B$/Er for \textit{H}$\parallel$ab, close to the 9.58 $\mu_B$ expected for the Er$^{3+}$ free ion. The fitted values of $\theta_p$ for \textit{H}$\parallel$c and \textit{H}$\parallel$ab are -0.34(4) and -22.5(2) K, respectively. The negative sign of $\theta_p$ suggests antiferromagnetic correlations in $\alpha$-ErAl$_3$, though crystal field level spitting may impact these results. 

\begin{figure}[t]
\centering
\includegraphics[width=0.49\textwidth]{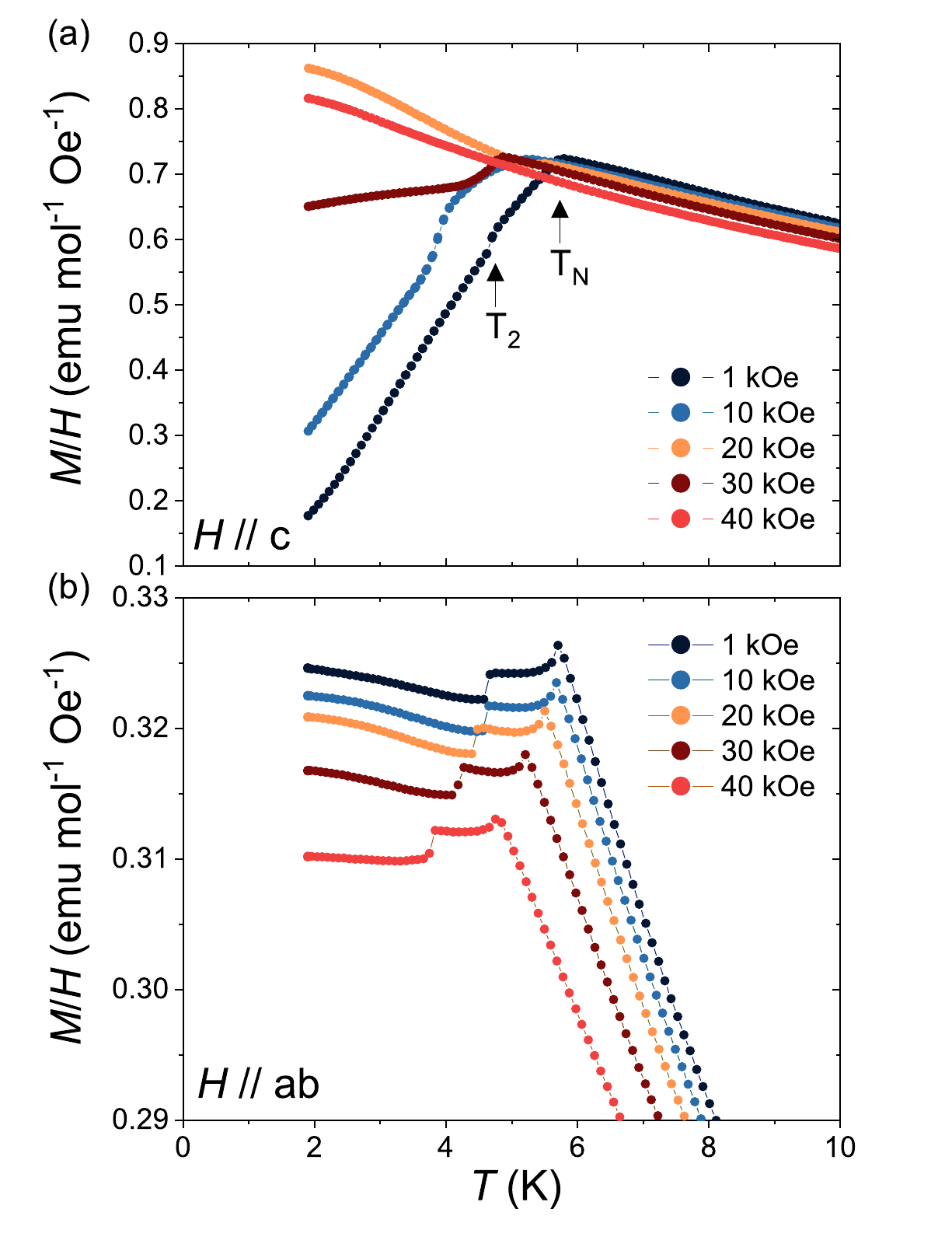}
\caption{Magnetic susceptibility versus temperature for various magnetic fields applied (a) parallel and (b) perpendicular to the \textit{c}-axis. Arrows indicate transitions at \textit{T$_N$} and \textit{T$_2$}.}
\label{fig:alpha_MvsT_fields}
\end{figure}

\begin{figure*}[ht!]
\centering
\includegraphics[width=0.85\textwidth]{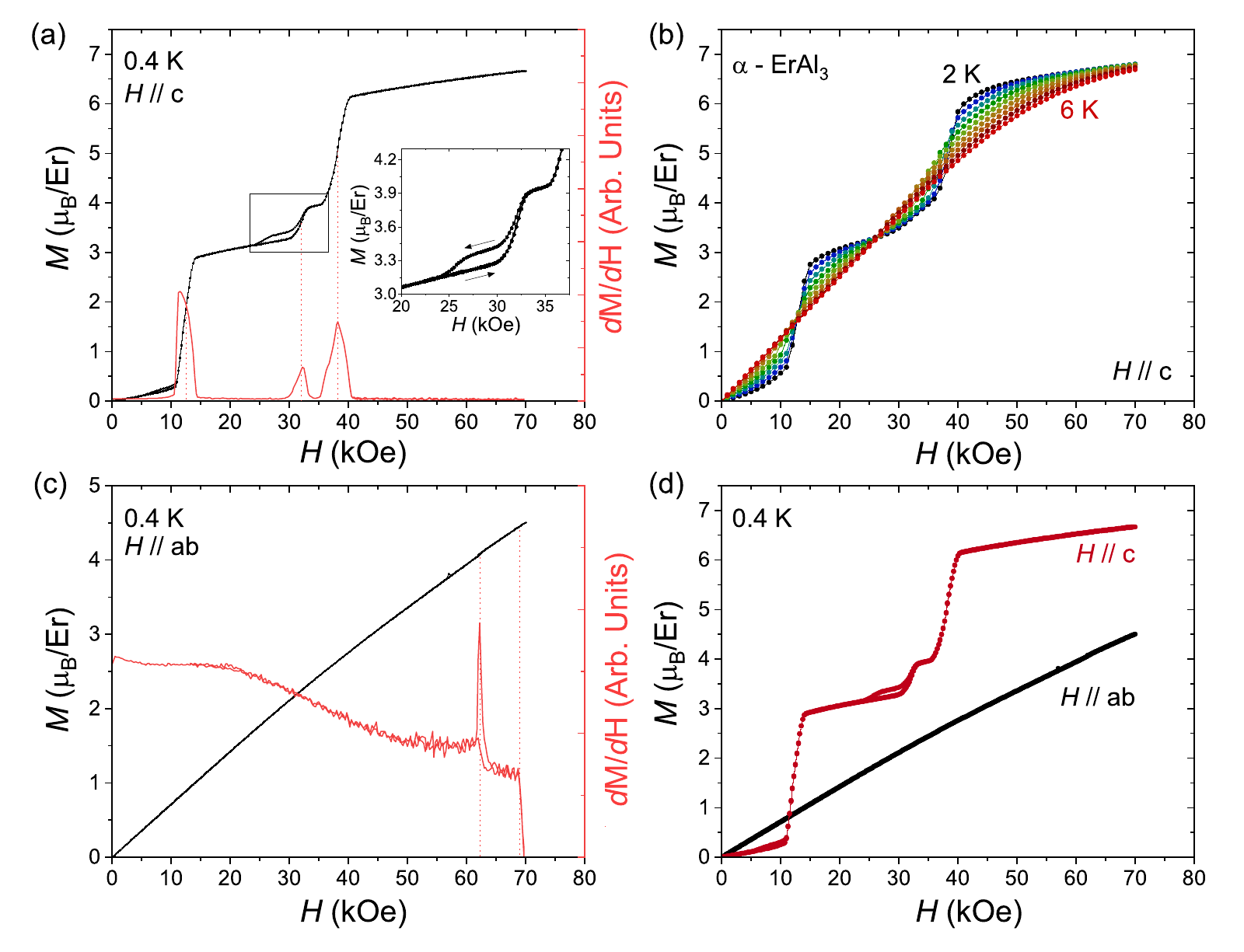}
\caption{(a) \textit{M}(\textit{H}) collected at 0.4 K for \textit{H}$\parallel$c together with the derivative \textit{dM}/\textit{dH}. The inset shows the zoom-in of the second field-induced transition. (b) Field-dependent magnetization at different temperatures ($\Delta$\textit{T} = 0.5 K) with field parallel to the \textit{c}-axis. (c) \textit{M}(\textit{H}) collected at 0.4 K for \textit{H}$\parallel$ab together with the derivative \textit{dM}/\textit{dH}. (d) Anisotropy of the magnetization  demonstrated by a comparison of the dataset at $T$ = 0.4 K.}
\label{fig:alpha_MvsH} 
\end{figure*}

Next we discuss the \textit{M}(\textit{T)} behavior near \textit{T}$_N$ for magnetic fields applied parallel (Fig.\ref{fig:alpha_MvsT_fields}a) and perpendicular (Fig.\ref{fig:alpha_MvsT_fields}b) to the \textit{c}-axis; for clarity, we show results for selected fields. The low-temperature magnetization data are characterized by a cusp at \textit{T$_N$} = 5.7 K and another transition at \textit{T$_2$} = 4.6 K is observed as a small kink. When \textit{H}$\parallel$ab, an increase in \textit{H} results in a continuous shift of \textit{T$_N$} and \textit{T$_2$} to lower temperatures.  However, the behavior of \textit{M}(\textit{T)} is more complicated when \textit{H}$\parallel$c. For fields below $\approx$ 12 kOe, the features at \textit{T$_N$} and \textit{T$_2$} are suppressed with increasing field. As the applied field is further increased, a qualitative change in the \textit{M}(\textit{T}) behavior is observed. For instance, as shown in Fig.\ref{fig:alpha_MvsT_fields}a, the magnetization rises upon cooling below the transitions for \textit{H} = 20 kOe, and the transition(s) has a very small effect on \textit{M}(\textit{T}). This slight increase in \textit{M} upon cooling through the transition may suggest a non-compensated (canted) antiferromagnetic spin structure in this \textit{H}-\textit{T} region of the phase diagram.  With a further increase to \textit{H} = 30 kOe, the magnetization is again characterized by a decrease upon cooling below the transitions. By $\approx$ 40 kOe, the transition is suppressed for \textit{H}$\parallel$c. For \textit{H}$\parallel$ab, a field larger than 66 kOe is required to suppress the transitions below 2 K. To construct the \textit{H}-\textit{T} phase diagram (discussed later), the values of the magnetic ordering temperatures were determined from the maximum of the derivatives \textit{d}(\textit{MT})/\textit{dT} for all applied fields \cite{fisher1962relation}.

To gain further information about the magnetic response, we measured the magnetic field dependence of the magnetization at various temperatures. Isothermal \textit{M}(\textit{H}) data with \textit{H} applied along the \textit{c}-axis and in the \textit{ab}-plane are presented in Fig.\ref{fig:alpha_MvsH}. There are several important features. First, the field-dependent magnetization of $\alpha$-ErAl$_3$ is characterized by well-defined metamagnetic transitions when a magnetic field is along the \textit{c}-axis (see Fig.\ref{fig:alpha_MvsH}a). At \textit{T} = 0.4 K, \textit{M}(\textit{H}) exhibits a step like behavior near 12 kOe and \textit{M} reaches a plateau of about 3.2 $\mu_B$/Er (note that this value is close to \textit{M}/\textit{M$_s$} = 1/2, where \textit{M$_s$} is the nearly-saturated magnetization above the last metamagnetic transition). The first metamagnetic transition has a small hysteresis in the magnetic field range of 6 to 11 kOe. For the second transition, occurring at approximately 32 kOe, we noted a broad, prominent hysteresis loop as manifested by the difference in \textit{M}(\textit{H}) for increasing and decreasing field measurements (see the inset of Fig.\ref{fig:alpha_MvsH}a). To investigate how this transition varies with temperature, we measured \textit{M}(\textit{H}) with a small temperature step size of $\Delta$\textit{T} = 0.1 K (not shown here) in the temperature range from 2 to 3 K. The data were used to construct the \textit{H}-\textit{T} phase diagram discussed later. For \textit{H} $>$ 50 kOe, the \textit{M}(\textit{H}) curve nearly saturates, reaching the value of \textit{M$_s$} $\approx$ 6.7 $\mu_B$/Er at 70 kOe and 0.4 K, which is lower than the value for free Er$^{3+}$ (9$\mu_B$). The value of the magnetization at 2 K and 135 kOe (not shown here) is \textit{M} $\approx$ 7.4$\mu_B$/Er, still below the full free ion moment and still increasing slightly with increasing field. This difference may be caused by the CEF effects, strong enough to decrease the saturation magnetization substantially. The rising \textit{M}(\textit{H}) could indicate a low-lying level, a higher-field metamagnetic transition, or prominent fluctuations. For higher temperatures (see panel (b) where \textit{M}(\textit{H}) curves are plotted for the temperature range 2 - 6 K), steps in \textit{M}(\textit{H}) become smeared, and the field-induced transitions vanish at temperatures close to \textit{T$_N$}.

When the magnetic field is applied in the \textit{ab}-plane, the magnetization has almost linear field dependence without indications of a metamagnetic transition. There are, however, slight changes in slope near \textit{H} = 62 kOe and \textit{H} = 68 kOe (see \textit{dM}/\textit{dH} plotted in Fig.\ref{fig:alpha_MvsH}c). We also note that the magnetization for \textit{H}$\parallel$ab is substantially smaller and does not saturate by 70 kOe (panel (d) in Fig. \ref{fig:alpha_MvsH}). Such a variation in magnetization in different directions indicates strong anisotropy brought about by CEF and/or anisotropic interactions. Moreover, \textit{M}(\textit{H}) data also indicate that the \textit{c}-axis is the magnetic easy axis, in agreement with the \textit{M}(\textit{T}) data. To construct a \textit{H}-\textit{T} phase diagram, we analyzed the derivatives \textit{dM}/\textit{dH} as a function of magnetic field at different temperatures (see \textit{dM}/\textit{dH} curves plotted for 0.4 K in Figs. \ref{fig:alpha_MvsH}a and \ref{fig:alpha_MvsH}c).

Our data demonstrate that $\alpha$-ErAl$_3$ undergoes successive transitions induced by a field applied along the easy-axis. Similar successive magnetization plateaus have been reported in many rare-earth compounds including some Er-based magnetic materials, e.g., the triangular-net Ising antiferromagnet ErGa$_2$ \cite{doukoure1982metamagnetism, kurumaji2024metamagnetic, onuki2023anomalous}, the quasi-one-dimensional Ising magnet TbTi$_3$Bi$_4$ with zigzag spin chain \cite{ortiz2024intricate, guo20241}, and in a Kagome ice compound HoAgGe with Ising spins lying in the kagome plane \cite{li2022low,zhao2020realization}. Further experiments are necessary to understand the origin of the successive magnetization plateaus in $\alpha$-ErAl$_3$, though the single ion anisotropy is likely an essential component of this physics.

\begin{figure}[t]
\centering
\includegraphics[width=0.49\textwidth]{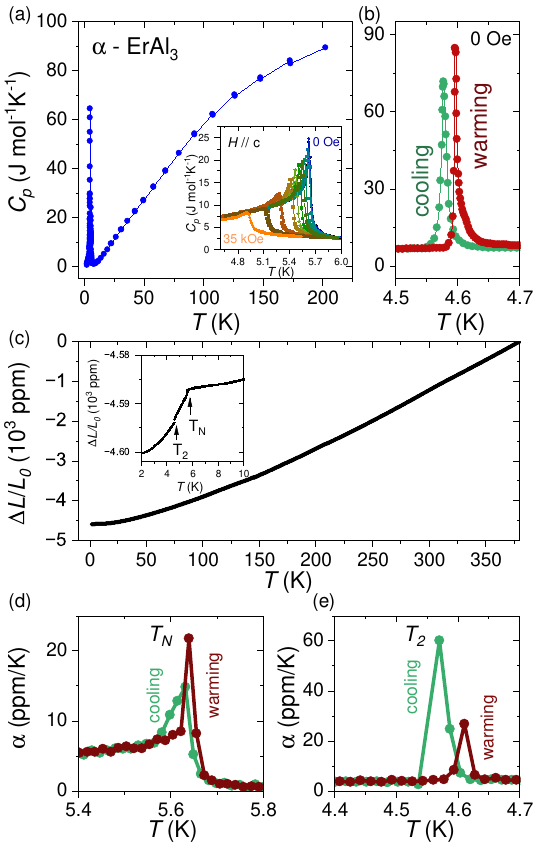}
\caption{(a) Temperature dependence of the heat capacity for $\alpha$-ErAl$_3$ single crystals in \textit{H} = 0 Oe. Inset shows the evolution of the AFM peak under applied magnetic fields up to 35 kOe. (b) The zero-field data obtained from the warming and cooling cycles in the vicinity of the transition at \textit{T$_2$}. (c) The relative contraction upon cooling, with the sample length \textit{L$_0$} obtained at 380 K. Inset depicts an expanded view at low temperatures. (d) and (e) The linear coefficient of thermal expansion $\alpha$(\textit{T}) for $\alpha$-ErAl$_3$, measured along the \textit{c}-axis in zero field.}
\label{fig:HC-alpha}
\end{figure}

Heat capacity measurements were carried out to further characterize the magnetic transitions. The temperature dependence of the zero-field heat capacity \textit{C$_p$}(\textit{T}) for $\alpha$-ErAl$_3$ is shown in Fig.\ref{fig:HC-alpha}a. At low temperatures, there are sharp anomalies near \textit{T$_N$} at 5.6 K and at \textit{T$_2$} = 4.6 K, confirming the bulk nature of magnetic transitions observed in magnetization measurements. The peak at \textit{T$_2$} is extremely sharp, with thermal hysteresis appearing in the warming and cooling cycle of \textit{C$_p$}(\textit{T}), as depicted in Fig.\ref{fig:HC-alpha}b. To account for this, in the vicinity of \textit{T$_2$}, the data were analyzed using a single-slope expression, which is an appropriate approach for first-order phase transitions. The first-order nature of this transition may be attributed to the crossover from commensurate magnetic order to an incommensurate or non-collinear magnetic state upon warming. Several rare-earth intermetallic compounds, such as TbNi$_2$Ge$_2$ \cite{islam1998neutron}, GdAuAl$_4$Ge$_2$ \cite{feng2022magnetic}, and SmAuAl$_4$Ge$_2$ \cite{feng2024complex} exhibit similar first-order transitions associated with changes in the magnetic structure.  Above the magnetic transition, we did not observe signatures of any Schottky anomalies that might be associated with the crystal fields levels, and the lack of an appropriate phonon-background makes it difficult to perform any additional analysis.

The low-temperature \textit{C$_p$}(\textit{T}) data measured at different applied magnetic fields with \textit{H}$\parallel$c are plotted in the inset of Fig.\ref{fig:HC-alpha}a. It is evident that \textit{T$_N$} shifts to lower temperatures and the maximum in the heat capacity decreases with increasing \textit{H} as expected. These data are used later to help construct the magnetic phase diagram in the \textit{H}-\textit{T} plane. To estimate \textit{T$_N$} at each field we take the temperature of the peak in \textit{C$_p$}(\textit{T}). 

\begin{figure}[b]
\centering
\includegraphics[width=0.49\textwidth]{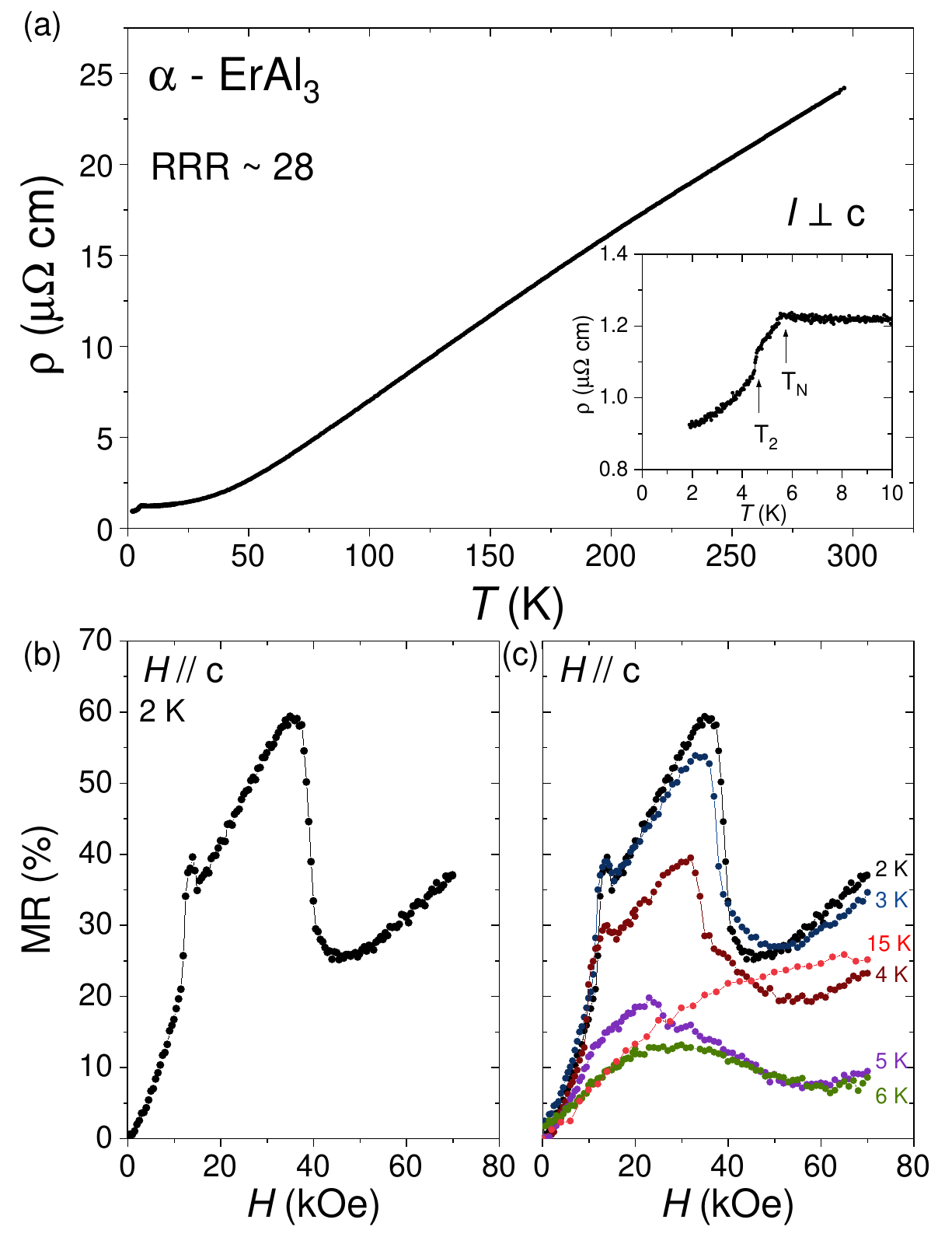}
\caption{(a) Temperature dependence of in-plane resistivity for $\alpha$-ErAl$_3$ single crystal in zero field. Inset shows the low-tempearture data around the magnetic transitions. (b) The magnetoresistafnce at 2 K for \textit{H}$\parallel$c. (c) Magnetoresistance with \textit{H}$\parallel$c at various temperatures.}
\label{fig:MR}
\end{figure}

Dilatometry was used to investigate the lattice response to the magnetic transitions in $\alpha$-ErAl$_3$. The relative length change $\Delta$\textit{L}/\textit{L$_0$} along [001] in zero magnetic field is plotted in Fig.\ref{fig:HC-alpha}c. The inset shows the dilation data at low temperatures, where anomalies are present at \textit{T$_N$} and \textit{T$_2$}. Through inspection of the enlarged view it is found that the \textit{c}-axis contracts rapidly while cooling through \textit{T$_N$}, which is associated with the onset of long-range magnetic order. Figs.\ref{fig:HC-alpha}d and \ref{fig:HC-alpha}e present the coefficient of thermal expansion $\alpha$(\textit{T}) = (1/\textit{L$_0$})(\textit{dL}/\textit{dT}) at low temperatures, with data on cooling-warming shown. The anomaly at \textit{T$_N$} is associated with a discontinuity or step in $\alpha$(\textit{T}) above/below the transition, a hallmark of a second-order type phase transition \cite{spachmann2021magnetoelastic,sakarya2003dilatometry,jaime2018magnetoelastics} (see panel (d)). In contrast, the anomaly at \textit{T$_2$} appears as a sharp peak with thermal hysteresis (panel (e)), confirming the first-order nature of this transition. Dilatometry, known for its high sensitivity to small lattice changes, strongly supports our magnetization and heat capacity measurements and confirms the existence of two consecutive phase transitions in zero field. Our investigation of the thermal expansion data for a $\alpha$-ErAl$_3$ single crystal do not reveal any additional anomalies. Previous studies on polycrystalline samples have reported a ferromagnetic transition occurring at higher temperatures \cite{ahmed2024large}, however, our results do not reveal any anomalies in the relative length change in those temperatures. It is important to highlight the fact that we were unable to examine the impact of the magnetic field on the thermal expansion due to the significant torque observed in the trigonal $\alpha$-ErAl$_3$.  

The temperature dependence of the electrical resistivity ($\rho$) measured within the \textit{ab}-plane of a $\alpha$-ErAl$_3$ single crystal is presented in Fig.\ref{fig:MR}a. The $\rho$(\textit{T}) exhibits a typical metallic behavior with the RRR [$\rho$(300 K)/$\rho$(2 K)] $\approx$ 28, indicating high quality single crystals. The resistivity decreases smoothly with decreasing \textit{T} until a sharp drop near the magnetic transition temperature. A two-step drop in the zero-field resistivity at \textit{T$_N$} and \textit{T$_2$} (see inset in Fig.\ref{fig:MR}a) is likely the result of substantial reduction in the spin-disorder scattering. 

The magnetoresistance (MR) of the $\alpha$-ErAl$_3$ single crystal up to \textit{T} = 6 K for \textit{H}$\parallel$c is shown in Fig.\ref{fig:MR}. The presented data were symmetrized using data from positive and negative applied fields. The MR is defined as [$\rho$(\textit{H})-$\rho$(0)]/$\rho$(0), where $\rho$(\textit{H}) and $\rho$(0) are the resistivities in the presence and absence of a magnetic field, respectively. At $T$ = 2 K (see Fig.\ref{fig:MR}b), the MR exhibits a kink near \textit{H} $\approx$ 13 kOe after the sharp rise to about 40\%. As the magnetic field is increased to 40 kOe, we observe a remarkable, sharp decrease in the MR associated with the third metamagnetic transition. When the magnetic field is greater than 50 kOe at 2 K, the magnetoresistance has a near-linear increase without saturation. Above 4 K (see Fig.\ref{fig:MR}c), the first anomaly disappears, and the second peak becomes broader and shifts to lower fields. At $T$ = 6 K, the MR still possesses a broad maximum, which might relate to the presence of strong spin-spin correlations in the vicinity near \textit{T$_N$}; a rather monotonic increase in MR is observed at higher temperatures. The anomalies observed in the MR data are well-aligned with the first and third metamagnetic transitions, indicating a coupling of the magnetism and the free carriers' lifetimes (scattering events) and/or band characteristics (density of states, velocities).  Further experiments are required to understand the microscopic origins. Large and positive MR containing anomalies near metamagnetic transitions have also been observed in similar rare earth intermetallics, such as HoAgGe with Ising-like spins arranged on a kagome lattice \cite{li2022low,zhao2020realization} and Pr$_3$MgBi$_5$ with a disordered kagome lattice \cite{han2023complex}.

\begin{figure}[t]
\centering
\includegraphics[width=0.49\textwidth]{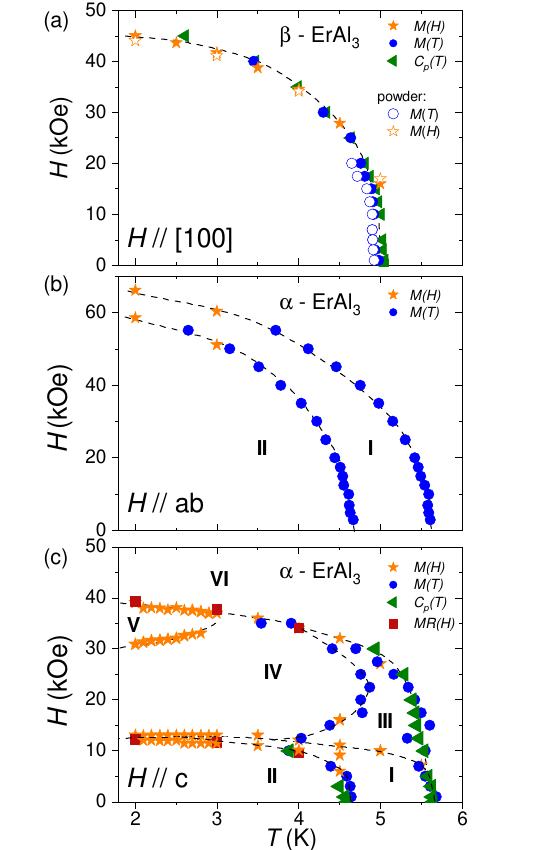}
\caption{\textit{H}-\textit{T} phase diagram of the (a) cubic $\beta$-ErAl$_3$, and the (b) trigonal $\alpha$-ErAl$_3$ when \textit{H}$\parallel$ab or (c) \textit{H}$\parallel$c. Lines are guide to the eye. For $\alpha$-ErAl$_3$, the phase transitions between I - II, IV - V appear to be discontinuous.}
\label{fig:diagram}
\end{figure}

Based on the experimental data presented above, we have constructed the \textit{H}-\textit{T} phase diagrams (see Fig.\ref{fig:diagram}) for the cubic $\beta$-ErAl$_3$ and the trigonal $\alpha$-ErAl$_3$ when \textit{H}$\parallel$c or \textit{H}$\parallel$ab. As shown in Fig.\ref{fig:diagram}a, the phase diagram for the cubic $\beta$-ErAl$_3$ comprises an AFM ordered phase below 5.1 K, which is fully suppressed under an applied magnetic field of 45 kOe. It is worth noting that the data points determined from magnetic measurements for crushed crystals (powder) are in agreement with the data obtained for the single crystal with \textit{H}$\parallel$[100].  The phase boundaries were determined based on several key observations, including the peak positions of derivatives of the magnetization data, the peaks in the \textit{C$_p$}(\textit{T}) data, and anomalies in the MR curves. The estimated error bars remain within the size of the data markers. Note that the phase boundaries gained from different experiments are consistent with each other.  

The phase diagram for the trigonal $\alpha$-ErAl$_3$ is substantially more complex than that of the cubic polymorph.  For \textit{H}$\parallel$ab (see Fig.\ref{fig:diagram}b), there are two phases that are both continually suppressed by increasing the applied field.  The phase diagram for \textit{H}$\parallel$c (see Fig.\ref{fig:diagram}c) is much more complex, with five magnetically ordered phases present below 40 kOe.  According to prior studies on the magnetic structure \cite{bargouth1971magnetic}, $\alpha$-ErAl$_3$ exhibits a commensurate stripe antiferromagnetic order with a  magnetic unit cell of dimensions 2\textit{a}, 2\textit{a}, \textit{c}. The zero-field neutron diffraction studies were conducted at 4.2 and 2.15 K, which are below the second transition temperature \textit{T$_2$} = 4.6 K, and thus the commensurate phase is the ground state spin structure (phase II). As discussed above, the transition between phases I and II is discontinuous, which might point to an incommensurate or non-collinear spin structure in phase I, though the microscopic nature of phase I has yet to be characterized. Based on M{\"o}ssbauer spectroscopy studies at 4.2 K \cite{weber1977magnetic}, Weber proposed that the ground state magnetic structure of $\alpha$-ErAl$_3$ consists of three different antiferromagnetic Er sublattices (1,2,3), with magnetic moments $\mu_1$ = (4.78 ± 0.16) $\mu_B$, $\mu_2$ = (5.97 ± 0.16) $\mu_B$, and $\mu_3$ = (7.10 ± 0.16) $\mu_B$. This likely relates to the nuclear (paramagnetic) structure, which contains three unique Er sites that are each in different layers along [001] (the layer types I, II, and III as in Fig. \ref{fig:crystal}).

Upon application of a magnetic field oriented along the $c$-axis, the magnetic transitions evolve in a complex manner to produce the phase diagram shown in Fig. \ref{fig:diagram}c. Consider first the ground state, where the applied field produces the sequential plateaus illustrated best at $T$ = 0.4 K in Fig. \ref{fig:alpha_MvsH}d. The first plateau corresponds to phase IV, while the small intermediate plateau corresponds to the pocket-like phase V. The presence of hysteresis at the boundary between these phases suggests that the transition from phase IV to phase V is of first order. The metamagnetic transition near 40 kOe leads to a nearly saturated state, likely a polarized paramagnetic state akin to ferromagnetism. The phase boundaries for phases IV and V share similar characteristics - when approaching the phase from the lower field boundary, increasing the applied field increases the stability of the phase (the field pulls the phase to higher $T$). Such behavior may be associated with a canted antiferromagnetic state (sometimes called a weak ferromagnet), which shares the characteristic of a ferromagnet where the applied field increases the apparent transition temperature by polarizing the moments. These behaviors correspond well with the temperature-dependent magnetization data, for instance the rise in $M$($T$) for $H$ = 20 kOe below the transition is suggestive of a non-compensated antiferromagnetic state (Fig. \ref{fig:alpha_MvsT_fields}a). On the upper side of these phase boundaries, the applied field decreases the transition temperature, thereby creating the pocket-like phase regions. The region where phases I through IV nearly intersect is difficult to map. Complementary experimental and theoretical investigations are necessary to fully elucidate the mechanisms underlying the transitions between the identified magnetic phases and to better understand the behavior in this portion of the diagram. Regardless, the existence of these various phases evidences a strong competition of different energies (competing interactions, local anisotropy, and perhaps lattice strain). Similar magnetic transitions and complex interactions have been observed in rare-earth systems such as GdAuAl$_4$Ge$_2$ \cite{feng2022magnetic}, SmAuAl$_4$Ge$_2$ \cite{feng2024complex}, DyRh$_2$Si$_2$ \cite{kliemt2023moment}, EuCd$_2$P$_2$ \cite{usachov2024magnetism} and Ce$_3$ScBi$_5$ \cite{xu2024magnetization}.

The complex behavior of $\alpha$-ErAl$_3$ is likely driven, in part, by the influence of long-range Ruderman-Kittel-Kasuya-Yosida (RKKY) interactions and significant local anisotropy induced by crystal fields. The anisotropy, either caused by Ising character of the moments or anisotropic interactions, probably drives the formation of the magnetic plateaus via spin-flipping processes. However, the existence of multiple plateaus complicates this simple picture and may indicate a rather complex spin structure for one of the intermediate phases (IV, V). The phases that do not display magnetization plateaus may be equally complex; for instance, the application of a magnetic field to a helical state is known to facilitate the formation of topologically nontrivial spin structures such as skyrmions. These considerations align with the scenario proposed by Chang and Mazin\cite{chang2024exploring}, which highlights how extended neighbor interactions and anisotropies can lead to similarly complex magnetic behavior and phase diagrams. Further work to understand the nature of these competing phases and the interactions that drive their formation is warranted.

\section{Conclusions}

In this study, we have investigated the magnetic, thermodynamic, and magnetotransport properties of cubic $\beta$-ErAl$_3$ and trigonal $\alpha$-ErAl$_3$ single crystals grown using Al flux. This study reveals the ability to grow single crystals of either polymorph based on the composition of the starting flux, with single crystals possessing resistivity ratios (RRR) greater than 20. A combination of physical property measurements establishes a magnetic transition at \textit{T$_N$} = 5.1 for $\beta$-ErAl$_3$. In $\beta$-ErAl$_3$, the application of a magnetic field continually suppresses \textit{T$_N$}.  The phase diagram for $\alpha$-ErAl$_3$ is more complex, especially for \textit{H}$\parallel$c. In zero field, the trigonal polymorph has transitions at \textit{T$_N$} = 5.7 K and  \textit{T$_2$} = 4.6 K. Thermal hysteresis in the anomalies of the specific heat and thermal expansion near \textit{T$_2$} evidence this to be a discontinuous phase transition. The ground state below \textit{T$_2$} is known to be a commensurate magnetic order with uniaxial anisotropy and in-plane antiferromagnetic couplings. The anisotropy leads to spin-flip like transitions when the field is applied along [001], and distinct magnetization plateaus are observed. The results are summarized in a phase diagram that reveals competition between five different magnetically ordered phases for \textit{H}$\parallel$c. Further studies to understand the nature and stabilization of these phases may help to generate a better understanding of Ising behaviors in rare earth intermetallics where RKKY interactions and local anisotropy play essential roles.

\section{Acknowledgment}
This work was supported by the U.S. Department of Energy, Office of Science, Basic Energy Sciences, Materials Sciences and Engineering Division.  We thank P. Park for useful discussions.

\bibliography{references}

\end{document}